\documentstyle[preprint,revtex,eqsecnum]{aps}
\begin{document}
\draft
\preprint{UAHEP-926}
\begin{title}
Statistical Mechanics of Black Holes
\end{title}
\author{B. Harms and Y. Leblanc}
\begin{instit}
Department of Physics and Astronomy, The University of Alabama,
\\ Box 870324, Tuscaloosa, Alabama 35487-0324
\end{instit}
\begin{abstract}
We analyze the statistical mechanics of a gas of neutral and
charged black holes.  The microcanonical ensemble is the only
possible approach to this system, and the equilibrium
configuration is the one for which most of the energy is carried
by a single black hole.  Schwarzschild black holes are found to
obey the statistical bootstrap condition.  In all cases,
the microcanonical
temperature is identical to the Hawking temperature of the most
massive black hole in the gas.
U(1) charges in general break the bootstrap property.
The problems of black hole decay and of quantum coherence are
also addressed.
\end{abstract}
\narrowtext

\section{INTRODUCTION}
\label{sec:level1}

It is by now well known that the area of a black hole event
horizon has received an interpretation in terms of
thermodynamical entropy and that the black hole mass has been
related, in this thermodynamical picture, to a temperature
called the Hawking temperature\cite{bek,hawk}.
Black holes should then
thermally evaporate, with the evaporation process ending when
the black hole reaches its extreme limit.

This state of affairs has been viewed quite recently with a
certain amount of criticism \cite{presk,cole,holz},
particularly in view of
the loss of quantum coherence in the black hole decay process.
The above interpretation of black hole phenomena such as the
decay (evaporation) process seems to imply that thermodynamics
is more fundamental than quantum mechanics in this problem, and
that pure states are converted into mixed states.  This state of
affairs has caused concern among researchers in this field.

Although such a problem originally arose in the mid-seventies, a
recent surge of interest has occurred in connection with the
discovery that black holes can carry quantum hair(for a nice
discussion see Ref.\cite{cole}).
Quantum hair
has the general effect of modifying the usual Hawking relation
for temperature in terms of mass and affecting the
expression for the entropy.  It was hoped that black holes would
carry quite a lot of quantum hair, enough to actually reduce the
thermal attributes down to zero and thereby generate the recovery
of quantum coherence.

Strings actually do carry quite a lot of quantum hair ( recall
the massive string excitations ) and so attempts
have been made to establish a connection between strings and
black holes\cite{thooft,verl}.

The degeneracy of string states at mass level m
is well known to be an
exponentially growing function of mass.  Typically, it is given
as,
\begin{eqnarray}
\rho_{string}(m) \sim c\; m^a e^{\beta_H m} \; ,
\end{eqnarray}
where the constants c, a and $\beta_H$ ( the inverse Hagedorn
temperature ) are model-dependent.  A microcanonical analysis
reveals that strings obey the following bootstrap condition
\cite{hage,fraut,carl},
\begin{eqnarray}
{\Omega (E) \over {\rho (E)}} \to 1 \; ;\ E \to \infty \; ,
\end{eqnarray}
where $\Omega$(E) is the microcanonical density of states.

Historically, the bootstrap condition was applied to the
statistical model of hadrons in an effort to explain the ever
increasing number of nuclear resonances found at higher energies
\cite{hage,fraut,carl}.

The degeneracy of states in mass space (Eq.(1.1)) was first
arrived at as a solution of the self-consistent constraint
(Eq.(1.2)), and only later was it recognized as a truly stringy
attribute.  In the context of the statistical model the
bootstrap constraint implied that hadronic resonances could be
viewed as being made of resonances, thereby replacing the
elementary particle concept.  Scattering theory was later
developed and the property of duality was further
demonstrated.  A by-product of the duality symmetry in the
scattering amplitudes is that the number of open channels in a
scattering process rises in parallel with the degeneracy of
states (Eq.(1.1)) as the energy is increased \cite{fraut}.

Like strings the degeneracy of states associated with a black
hole increases exponentially, with the argument of the
exponential now quadratic in mass, at least to leading order in
a U(1) charge expansion.

As an example, in natural units $(\hbar = c= G = 1)$, a
Schwarzschild ( neutral ) black hole has the following density
of states in mass space,
\begin{eqnarray}
\rho_{Schw} (m) \sim c\; e^{4\pi m^2}
\end{eqnarray}
where m is the mass of the black hole.  This result is a
non-perturbative quantum effect in the sense that it is obtained
from the WKB method and
that the argument of the exponential is of
order $\hbar^{-1}$(For a detailed derivation see
Ref.\cite{cole}.).
The constant c here represents the unknown
effects from the purely perturbative quantum field theoretical
sector of the theory.

The above density of states has been
given the interpretation \cite{bek,hawk} of a
thermodynamical system with entropy,
\begin{eqnarray}
s(m) = 4\pi m^2
\end{eqnarray}
and inverse temperature,
\begin{eqnarray}
\beta_{Hawking} (m) = {ds \over dm} = 8\pi m\; .
\end{eqnarray}

Eq. (1.5) however yields a negative specific heat,
\begin{eqnarray}
(C_V)_{Hawking}
= -\beta^2 {dm \over {d\beta}} = -{\beta^2_{Hawking}
\over {8\pi}}\; .
\end{eqnarray}

It is important to recall that the results in Eqs.
(1.4) - (1.6) can be
derived ( e.g. see Ref. \cite{cole} ) by analytically continuing
the expression for the black hole metric to imaginary time and
so to Euclidean space-time.  The temperature emerges as the
inverse period of the compact Euclidean time and its relation to
the black hole mass is determined from the requirement of the
absence of conical singularities of the Euclidean space-time.
This Matsubara-type method produces results which, in the
thermodynamical interpretation,
belong to the
canonical ensemble.

Although usually interpreted as a sign of instability, the
negative specific heat (Eq.(1.6)) represents nevertheless a
flaw in the above thermal interpretation.  The main
point here is that, within the canonical (thermal) ensemble the
specific heat is always a positive definite quantity
\cite{carl,thir}.  Gravitational systems (e.g. supernovae and
galaxies) are known to exhibit a negative microcanonical
specific heat \cite{thir}.  However, in the canonical ensemble
one sees only a phase transition, so that the canonical specific
heat is always positive.
The above
results therefore point in our judgment
to an inconsistency of the
thermal interpretation, not just an instability.

The present work is based on the observation that two
interpretations of the density of states (Eq.(1.3)) exist for
the
Schwarzschild black hole (or any black hole),
one of which of course must be wrong.
The above thermodynamical interpretation \cite{bek,hawk} is
one such interpretation.   However, as was pointed out, it leads
to undesirable features such as a negative canonical specific
heat as well as a breakdown of the laws of quantum mechanics as
pure states evolve into mixed states during the process of
black hole evaporation.  This interpretation seems to run into
problems with both thermodynamics and quantum mechanics.

In the other interpretation, we simply regard Eq. (1.3) as
the degeneracy of states of a quantum Schwarzschild black hole
at mass level m, in a way analogous to the degeneracy of states
in string theory.  In this
way the laws of quantum mechanics remain untouched and the
process of black hole evaporation can be understood from an
S-matrix theory point of view.

One then naturally becomes interested in the problem of
understanding the statistical mechanics of a gas of such
objects.

In the following sections we shall find that the microcanonical
ensemble is the unique sensible framework for analyzing this
problem.  At least for the Schwarzschild case it will be found
that the bootstrap constraint  (Eq.(1.2)) is met, hence
providing us with the novel view of a black hole as an object
itself
made of other
black holes, very much as in the old statistical model
of hadrons.

The equilibrium state of such a system is not thermal.  A
quantum coherent view of black hole decay (evaporation) can also
be obtained in a way analogous to strings.  Black hole states
decay into other black hole states.  Although the problem of
constructing black hole scattering amplitudes is not addressed
in this paper, we shall nevertheless demonstrate that the number
of open channels for n-body decay of a Schwarzschild black hole
does indeed grow precisely in parallel with density of states
(Eq.(1.3)), thereby allowing duality symmetry for black hole
scattering amplitudes.  Black holes may belong to a certain
class of string theories, as conjectured previously
\cite{thooft}.

Obvious applications of the considerations presented in this
paper lie in cosmology and the very early universe, as well as
in galaxy formation and the general theory of gravitational
collapse.

\section{A SIMPLE CASE: THE SCHWARZSCHILD BLACK HOLE}
\label{sec:level2}

In this section we analyze the statistical mechanics of a gas of
Schwarzschild black holes with degeneracy given by Eq.(1.3).  In
this problem as a working hypothesis we shall assume that the
equilibrium state (if any) can be achieved on time scales less
than the individual lifetimes of the black holes in the gas.
This is tantamount to neglecting the decay rates (hence
collision processes) so that one remains conveniently in the
ideal gas approximation.  There is a simple argument
establishing the non-existence of the canonical ensemble
description of a gas of black holes with degeneracy given by
Eq.(1.3).  Recalling the form (Eq.(1.1)) for strings, it is well
known that the canonical ensemble breaks down whenever the
exponential factor in Eq.(1.1) wins over the Boltzmann factor
$e^{-\beta m}$ in the statistical sum (integral over mass).  For
strings this occurs at temperatures above the Hagedorn
temperature $\beta_H^{-1}$.  For black holes however the
exponential factor always dominates and so the canonical
partition function diverges for all temperatures.  One may then
expect large disparities, which show as unbounded fluctuations
in the thermal ensemble, in the energy distribution among the
components of the gas.  This is in fact what happens.

We now turn to the unique approach to this problem, namely the
microcanonical description.  The microcanonical density of states
is written as follows,
\begin{eqnarray}
\Omega (E,V) = \sum_{n=1}^{\infty} \Omega_n (E,V)\; ,
\end{eqnarray}
in which we have
defined the density of states for the configuration
with n black holes as,
\begin{eqnarray}
\Omega_n (E,V) &=& \left[{V \over {(2\pi )^3}}\right]^n {1\over
{n!}}\prod_{i=1}^n \left\{ \int_{m_0}^{\infty} dm_i\;
\rho_{B.H.} (m_i) \int_{-\infty}^{\infty} d^3 p_i \right\}
\nonumber \\ & & \times \delta (E-\sum_{i=1}^n E_i )\;
\delta^3 (\sum_{i=1}^n \vec{p}_i)\; ,
\end{eqnarray}
where E is the total energy of the system and where V is the
volume of the gas.

We can write the product of the degeneracy of states as,
\begin{eqnarray}
\prod_{i=1}^n e^{a\; m_i^2} = e^{a\sum_{i=1}^n E_i^2}
e^{-a\sum_{i=1}^n \vec{p}_i^{\; 2}}\; ,
\end{eqnarray}
in which we have made the working assumption that black holes
obey the particle-like dispersion relation,
\begin{eqnarray}
E^2 = \vec{p}^{\; 2} +m^2 \; .
\end{eqnarray}

As is clear from Eq.(2.3), the high momentum states contribute
negligibly (they are actually exponentially suppressed) to the
microcanonical density of states.  Hence at high energy the
dominant contributions originate from mass.  Following Frautschi
and neglecting the momentum conservation $\delta$ - function in
Eq.(2.2), the momentum integrations simply become n decoupled
Gaussian integrals, the values of which can be absorbed into a
redefinition of the volume factor.  Eq.(2.2) then reduces to,
\begin{eqnarray}
\Omega_n (E,V) \simeq \left[
{V \over{(2\pi )^3}}\right]^n {1 \over {n!}}
\prod_{i=1}^n \int_{m_0}^{\infty} dE_i\; \rho_{B.H.} (E_i)\;
\delta (\sum_{i=1}^n E_i - E)\; .
\end{eqnarray}
Again, Frautschi has made a general analysis of systems with
degeneracy $\rho$(m) of the following generic form,
\begin{eqnarray}
\rho (m) = f(m)\; e^{b\; m^p}\; ,
\end{eqnarray}
where f(m) is a polynomial in m.  Substituting this form into
Eq.(2.5), we find the maximum value of the integrand to be
at $E_i = E/n$ for any p (i.e. the total energy is evenly
distributed among all parts).  For p $<$ 1 this is the dominant
configuration.  However for p $>$ 1 contributions from the
integration boundaries yield the dominant configuration in which
most of the energy is carried by a single black hole.  Such a
configuration is the one for which, e.g., the nth black hole
carries the energy $E_n = E-(n-1)m_0$ while the n-1 other black
holes carry energies $E_i = m_0$ (i =1,...,n-1).  Since a
Schwarzschild black hole corresponds to p = 2, the density of
states (Eq.(2.5)) finally becomes,
\begin{eqnarray}
\Omega_n (E,V) \simeq \left[
{cV \over{(2\pi )^3}}\right]^n {1 \over{n!}}
e^{4\pi [E-(n-1)m_0]^2} e^{4\pi (n-1)m_0^2}\; ,
\end{eqnarray}
an expression valid at high energy E.

The most probable equilibrium configuration is the one
satisfying the condition,
\begin{eqnarray}
{d\Omega_n (E,V) \over {dn}}\mid_{n=N(E,V)} \; = 0 \; .
\end{eqnarray}
We find,
\begin{eqnarray}
exp\Psi (N+1) = {cV \over {(2\pi )^3}} exp[s(m_0)-m_0
\beta_{Hawking}(E-(N-1)m_0)]\;,
\end{eqnarray}
Where s(x) is the Hawking entropy (Eq.(1.4)),
$\beta_{Hawking}(x) \equiv {ds(x) \over{dx}}$
and $\Psi (x)$ is the
psi function.

Now since the lightest object in the gas is the extreme
Schwarzschild black hole, we have $m_0 = 0$.  Therefore,
\begin{eqnarray}
exp\Psi (N+1) = {cV \over{(2\pi )^3}}\; .
\end{eqnarray}
The total entropy of the system is now given as,
\begin{eqnarray}
S(E,V) &\equiv &\ln\Omega (E,V) \simeq \ln\Omega_N (E,V)\nonumber
\\ &\simeq & N\ln \left[{cV \over{(2\pi )^3}}\right]
-\ln\Gamma (N+1) + s(E)
\end{eqnarray}
where N(V) is given by Eq.(2.10).  The inverse temperature
$\beta$ is obtained from the total entropy according to,
\begin{eqnarray}
\beta = {dS\over{dE}} = {\partial S\over{\partial N}}{\partial N
\over{\partial E}} + {\partial S\over{\partial E}}\; .
\end{eqnarray}
The first term on the far right is zero at the maximum value of
$\Omega$ (recall Eqs.(2.8),(2.11)).
We find quite generally that,
\begin{eqnarray}
\beta &=& {dS \over {dE}} = {ds \over {dE}} = \beta_{Hawking}
\nonumber \\ &=& 8\pi E\; .
\end{eqnarray}
So the microcanonical temperature for a gas of Schwarzschild
black holes is the same as the Hawking temperature of the black
hole carrying the greatest amount of energy in the gas.

The microcanonical specific heat is likewise negative,
\begin{eqnarray}
C_V = -\beta^2{dE \over{d\beta}} ={-\beta^2 \over{8\pi}} \; ,
\end{eqnarray}
a result identical to Eq.(1.6).  Although this result implies
that instabilities will develop if the gas is brought in contact
with a heat bath, it is not an inconsistent finding.  The
specific heat is allowed, in principle, to be negative in the
microcanonical ensemble.

{}From the formula,
\begin{eqnarray}
\beta P = {\partial S \over{\partial V}}\; ,
\end{eqnarray}
where P is the pressure, we obtain the following equation of
state,
\begin{eqnarray}
\beta P = {N \over V} \; ,
\end{eqnarray}
which is that of an ideal gas.

Finally, it is readily seen that the bootstrap condition is
satisfied for a gas of Schwarzschild black holes, provided (cf.
Eqs.(1.2), (1.3) and (2.11)),
\begin{eqnarray}
\left[{cV \over{(2\pi )^3}}\right]^N
{1\over{\Gamma (N+1)}} = c \; .
\end{eqnarray}
That this is so originates from the fact that an extreme
Schwarzschild black hole is massless.  Eq.(2.17) together with
Eq.(2.10) tells us that the size of the quantum corrections to
the density of states Eq.(1.3) determines the volume of the gas
in units of the Planck volume, as well as the size of the most
probable configuration.  Assuming N $\gg$ 1 and c $\gg$ 1,
one gets,
\begin{eqnarray}
V \sim {(2\pi )^3\over{c}}\ln c \; ; \ N \sim \ln c \; .
\end{eqnarray}

In the following sections we extend our analysis to gases of
black holes carrying U(1) electric charge, namely the
Reissner-Nordstr\"{o}m and dilaton black holes.

\section{THE REISSNER-NORDSTR\"{O}M BLACK HOLE}
\label{sec:level3}

In this section we analyze the case of a gas of charged
Reissner-Nordstr\"{o}m
black holes with identical individual charges
Q.  These individual charges are taken to be
much smaller than the total
energy E of the system.  For such a case the density of states
for the configuration with n black holes is given by the
following generalization of Eq.(2.7),
\begin{eqnarray}
\Omega_n (E,V,Q) \sim \left[{cV \over{(2\pi )^3}}\right]^n
{1\over{n!}}
e^{s(E-(n-1)m_0, Q)} e^{(n-1)s(m_0, Q)} \; ,
\end{eqnarray}
which is valid at high energy E, and where $s(m,Q)$
is the Hawking
entropy of a single Reissner-Nordstrom black hole of mass m,
\begin{eqnarray}
s(m,Q) = \pi m^2 \left[1+\sqrt{1 - {Q^2\over {m^2}}\; }\;
\right]^2\; .
\end{eqnarray}

The corresponding Hawking (inverse) temperature is given by the
relation,
\begin{eqnarray}
\beta_{Hawking} (m,Q) = {ds \over {dm}} = 2\pi m\left[1+
\sqrt{1-{Q^2
\over{m^2}}\; }\; \right]^2(1-{Q^2\over{m^2}})^{-1/2}\; .
\end{eqnarray}
The degeneracy of states for such a hole is given by,
\begin{eqnarray}
\rho_{RN}(m,Q) = c\; exp\; s(m,Q) \; .
\end{eqnarray}
Again for this case the dominant equilibrium configuration of
the gas is not thermal, but given rather by the state with one
very massive black hole and (n-1) light ones with mass $m_0$.

Since extreme Reissner-Nordstr\"{o}m black holes have mass m = Q,
and since these are the lightest elements of the gas, we have
the identification,
\begin{eqnarray}
m_0 = Q
\end{eqnarray}

The most probable configuration N(E,V,Q) is the one which
maximizes the density of states (Eq.(3.1)).  We find
\begin{eqnarray}
e^{\Psi (N+1)} = {cV \over {(2\pi )^3}} e^{s(Q,Q) -
Q\beta_{Hawking} (E-(N-1)Q, Q)}\; .
\end{eqnarray}
Now since E $\gg$ (N-1)Q, we get the following approximate
relation,
\begin{eqnarray}
\Psi (N+1) \sim \ln \left[{cV \over{(2\pi )^3}}\right]
- 8\pi Q E + O(Q^2)\;.
\end{eqnarray}
For large N the condition
\begin{eqnarray}
{cV\over{(2\pi )^3}} \gg e^{8\pi QE}\;,
\end{eqnarray}
must be satisfied.  Clearly at high enough
energy the above condition
cannot be met.  The most probable configuration at high energy
is the one for which N is as small as possible, i.e. N = 1.  In
the statistical bootstrap model of hadrons the configuration N =
1 corresponds to an elementary particle and is usually ruled
out.  In the present microcanonical formulation there is no
logical argument to exclude such a case.

The most probable equilibrium configuration of a gas of
Reissner-Nordstr\"{o}m black holes is then described by the
conditions,
\begin{eqnarray}
(N-1)Q &\ll &E \ll E_c \; ; \ N \gg 1 \nonumber \\
E &=& E_c \; ;\  N = 1\; ,
\end{eqnarray}
where $E_c$ is an ``ionization point'' determined by the
following formula,
\begin{eqnarray}
Q\beta_{Hawking} (E_c,Q) = s(Q,Q)+\ln \left[{cV\over {(2\pi
)^3}}\right] -\Psi (2) \; .
\end{eqnarray}
For small charge Q we get,
\begin{eqnarray}
E_c \sim {1\over{8\pi Q}} \left\{ \ln \left[
{cV\over{(2\pi )^3}}\right] -
\Psi (2)\right\} \; .
\end{eqnarray}

The total entropy of the gas is given as follows,
\begin{eqnarray}
S(E,V,Q) &\simeq& \ln \Omega_N (E,V,Q)\nonumber \\
& = & N\ln \left[{cV\over{(2\pi )^3}}\right] - \ln \Gamma
(N+1)\nonumber \\
& & + s(E-(N-1)Q,Q) + (N-1)\; s(Q,Q)\; ,
\end{eqnarray}
As in the Schwarzschild case the microcanonical temperature is
the same as the Hawking temperature of
the most massive black hole in the gas,
\begin{eqnarray}
\beta (E,V,Q) = \beta_{Hawking} (E-(N-1)Q,Q)\; ,
\end{eqnarray}
with N(E,V,Q) given by Eq.(3.6).  The equation of state is that
of an ideal gas,
\begin{eqnarray}
\beta P = {N\over V}
\end{eqnarray}

Clearly the bootstrap constraint cannot be met for $Q \neq 0$
and/or $N \neq 1$.  At the ``ionization point'', however, the
bootstrap constraint is trivially met (there is a single black
hole).  Inserting $N = 1$ and $E = E_c$ into Eq.(3.12) leads to
the volume contraint,
\begin{eqnarray}
V = V_c \equiv (2\pi )^3
\end{eqnarray}
The lone black hole occupies a region of space with a size of
the order of the Planck volume.

Inserting Eq.(3.15) into expression (3.11) for $E_c$ we find,
\begin{eqnarray}
E_c \sim {1\over{8\pi Q}}[\ln c - \Psi (2)]\; .
\end{eqnarray}
Of course, consistency requires that $\ln c > \Psi (2)$ .

For $E > E_c$ there is no equilibrium configuration.  It seems
plausible to speculate that some kind of phase transition may
occur at the ``ionization point''.

We close this section by providing approximate formulae in the
large N ($E \ll E_c$) region.
For large N Eq.(3.7) becomes,
\begin{eqnarray}
N \simeq {cV \over{(2\pi )^3}} e^{-8\pi QE}.
\end{eqnarray}
In this approximation the total entropy (Eq.(3.12)) becomes
\begin{eqnarray}
S
&\simeq& 4\pi E^2 + 8\pi QE + {cV \over{(2\pi )^3}} e^{-8\pi QE}
+ O(Q^2)
\end{eqnarray}
This expression gives for $\beta$,
\FL
\begin{eqnarray}
\beta \simeq  8\pi E + 8\pi Q \left[ 1
-{cV \over {(2\pi )^3}} e^{-8 \pi QE}\right] + O(Q^2) \; ,
\end{eqnarray}
which is, to the same approximation, the inverse Hawking
temperature.  In this case again
the specific heat is negative.
\begin{eqnarray}
C_V &=& -\beta^2 \left( {d\beta \over{dE}}\right)^{-1}
\nonumber \\
&\simeq& -8\pi [E^2 + 2QE\;(
1-{cV\over{(2\pi )^3}}e^{-8\pi QE})] +
O(Q^2)\; .
\end{eqnarray}
Thus the charged black hole, like the Schwarzschild black
hole, cannot reach thermal equilibrium with its surroundings.

In the next section we analyze a somewhat more general case, the
so-called dilaton black hole.

\bigskip \medskip
\section{THE DILATON BLACK HOLE}
\label{sec:level4}

The dilaton black hole \cite{gib,garf} is somewhat similar to the
Reissner-Nordstr\"{o}m
black hole with the added complexity of the
effects of an additional dilaton field coupling.

The microcanonical analysis of a gas of dilaton black holes is
identical to the case of the Reissner-Nordstr\"{o}m black hole
gas.
The Hawking entropy and temperature however are now given by,
\begin{eqnarray}
s(m,Q) = \pi m^2 \left[
1+\sqrt{1-{(1-a^2)Q^2\over{m^2}}\; }\; \right]^2 \left( 1 -
{(1+a^2)Q^2
\over{m^2[1+\sqrt{1-{(1-a^2)Q^2\over{m^2}}}]^2}}\right)^{{2a^2
\over{1+a^2}}}
\end{eqnarray}
and
\begin{eqnarray}
\beta_{Hawking} (m,Q) = 4\pi m\left[
1+\sqrt{1-{(1-a^2)Q^2\over{m^2}}\; }\; \right]
\left(1-{(1+a^2)Q^2\over{m^2[1
+\sqrt{1-{(1-a^2)Q^2\over{m^2}}}]^2}}\right)^{{a^2-1 \over {a^2
+1}}}\; .
\end{eqnarray}
Again the degeneracy of states is given as,
\begin{eqnarray}
\rho (m,Q) = c\;  exp [s(m,Q)]\; .
\end{eqnarray}
Notice that the case a = 0 reduces to the Reissner-Nordstr\"{o}m
black hole.

The extreme dilaton black hole has a mass,
\begin{eqnarray}
m_0^2 ={Q^2 \over{1+a^2}} \; .
\end{eqnarray}
Repeating the analysis of the previous section, the ``ionization
point'' (N = 1) at high energy $E_c$ is found to be given by the
formula,
\begin{eqnarray}
{Q\over{\sqrt{1+a^2}}} \beta_{Hawking}
(E_c,Q) = s({Q\over{\sqrt{1+a^2}}},Q) +
\ln \left[{cV\over{(2\pi )^3}}\right] - \Psi(2)\;.
\end{eqnarray}
Again, for small Q, the result is found to be similar to the
R-N case,
\begin{eqnarray}
E_c \sim {\sqrt{1+a^2}\over{8\pi Q}}\left\{ \ln \left[
{cV\over{(2\pi
)^3}}\right] - \Psi (2)\right\} \; .
\end{eqnarray}

For large N the density of states $\Omega$ is a maximum at,
\begin{eqnarray}
N \simeq {cV \over {(2\pi )^3}} e^{-8\pi QE /(1+a^2)^{1/2}}\; ;
\ ({(N-1)Q\over{\sqrt{1+a^2}}} \ll E \ll E_c)\; .
\end{eqnarray}

The total entropy of the gas is
\begin{eqnarray}
S(E,V,Q) &\simeq& \ln \Omega_N \nonumber \\
&\simeq&
N \ln \left[ {cV \over{(2\pi )^3}}\right] - \ln \Gamma (N+1)
+ s(E-{(N-1)Q\over{\sqrt{1+a^2}}}, Q) \nonumber \\
& & + (N-1)\;
s({Q\over{\sqrt{1+a^2}}},Q)\; .
\end{eqnarray}
The microcanonical temperature is again easily derived.  We
find,
\begin{eqnarray}
\beta(E,V,Q) &=& {dS\over{dE}} = {\partial S\over{\partial
N}}{\partial N\over{\partial E}} + {\partial S\over{\partial E}}
\nonumber \\
&=& {\partial s(E-{(N-1)Q\over{\sqrt{1+a^2}}},Q)\over{\partial
E}} = \beta_{Hawking} (E-{(N-1)Q\over{\sqrt{1+a^2}}},Q)\; ,
\end{eqnarray}
in which we once again used the fact that N(E,V,Q) is the most
probable configuration, i.e. ${\partial S\over{\partial N}} =
0$.

For large N we have the following approximate expression
\begin{eqnarray}
\beta &\simeq& 8\pi E + {8\pi Q\over{\sqrt{1+a^2}}}\left[ 1 -
{cV\over{(2\pi )^3}} e^{-{8\pi QE\over{\sqrt{1+a^2}}}}\right] +
O({Q^2\over{\sqrt{1+a^2}}})\; ;\nonumber \\
& & \left({(N-1)Q\over{\sqrt{1+a^2}}} \ll E \ll E_c\right)\; .
\end{eqnarray}
Finally, the microcanonical specific heat is given by
\begin{eqnarray}
C_V = -\beta^2\left({d\beta\over{dE}}\right)^{-1} =
-\beta^2\left[{\partial\beta\over{\partial E}} +
{\partial\beta\over{\partial N}}{\partial N\over{\partial
E}}\right]^{-1}\; .
\end{eqnarray}
Recalling Eq.(4.9), we find,
\begin{eqnarray}
C_V &=& -\beta^2\left[{\partial\beta \over{\partial E}} +
{\partial\over{\partial E}}\left({\partial s\over{\partial
N}}\right){\partial N\over{\partial E}}\right]^{-1}\nonumber \\
&=& -\beta^2\left({\partial\beta\over{\partial
E}}\right)^{-1}\left[ 1-{Q\over{\sqrt{1+a^2}}}{\partial
N\over{\partial E}}\right]^{-1}\; .
\end{eqnarray}
Therefore,
\begin{eqnarray}
C_V (E,V,Q) = (C_V)_{Hawking}
(E-{(N-1)Q\over{\sqrt{1+a^2}}},Q)\left[1-{Q\over{\sqrt{1+a^2}}}
{\partial N\over{\partial E}}\right]^{-1}\; .
\end{eqnarray}

Recalling Eq.(4.7) one finds that ${\partial N\over{\partial E}}
< 0$  so that the sign of the microcanonical specific heat is
determined by that of the Hawking specific heat.

Although ${\partial N\over{\partial E}} < 0$
is formally valid at high energy, it is
tempting and probably valid
to extrapolate to the low energy domain where the total
energy approaches the extreme limit $E \to
{NQ\over{\sqrt{1+a^2}}}$ of the gas with
total charge NQ.
It is known that, for $a < 1$, the dilaton
black hole has a positive Hawking specific heat as it approaches
its extreme limit,  whereas for $a > 1$ the specific heat is
negative\cite{holz}.
It may be an interesting problem to analyze the
implications of such properties from the viewpoint of physical
processes in the very early universe.
\vfill \eject
\section{QUANTUM COHERENT BLACK HOLE DECAY}
\label{sec:level5}

The previous sections were devoted to the statistical mechanics
of black holes.  In this section we attempt to uncover a few
facts about (quantum coherent) black hole decay and scattering.

A property particular to strings and dual models is the duality
of the scattering amplitudes.  The 4-point amplitude for
instance can be expressed as a sum over resonances either in the
s- or t-channel, even at very high energy.  Frautschi
\cite{fraut}
has pointed
out that in order for duality to hold the number of n-body
channels open in the statistical model of hadrons, and so the
total number of open channels, must rise precisely in parallel
with the number of resonances as the center of mass energy is
increased,
\begin{eqnarray}
N_n (m) \sim \rho_{string} (m) ; \ m \to \infty \; ,
\end{eqnarray}
where $N_n (m)$ is the number of open n-body channels at center
of mass energy m.  An explicit expression for the 2-body case is
given by,
\begin{eqnarray}
N_2 (m) = {1\over {2!}}\int_{m_0}^{m-m_0}dm_2\; \rho (m_2)
\int_{m_0}^{m-m_2} dm_1\; \rho (m_1)\; .
\end{eqnarray}
Consequently, if duality can be argued to be
a symmetry of, say, Schwarzschild
black hole scattering amplitudes, one should be able to support
it by a direct calculation of Eq.(5.2), making use of the
degeneracy (Eq.(1.3)).  Inserting Eq.(1.3) into Eq.(5.2), we
obtain ($m_0 = 0$),
\begin{eqnarray}
N_2 (m) = {c^2\over{2!}}\int_0^m dm_2 \int_0^{m-m_2} dm_1\;
e^{4\pi (m_1^2 + m_2^2)} \; .
\end{eqnarray}
Clearly, contributions from the region $m_2 \sim m$ give
negligible results.  The dominant contribution is obtained when
$m \gg m_2$ and $m_1 \sim m$.  This is the same situation which
occurred in the evaluation of the density of states with n black
holes, in which most of the energy was carried by a single black
hole and the n-1 others shared the tiny remnants.  In this
approximation we obtain,
\begin{eqnarray}
N_2 (m) &\simeq &{c^2\over{2}} e^{4\pi m^2} \nonumber \\
&=& {c\over {2}}\; \rho_{Schw} (m)\; ; \ m\to \infty \; .
\end{eqnarray}
This argument can be extended to any n, and one finds the same
result, namely that as in the string theory case, the number of
open channels does grow in parallel with the degeneracy of states
as energy is increased.

It is now very plausible to argue that the black hole (at least
the Schwarzschild black hole) scattering amplitudes do share the
property of duality with the string scattering amplitudes.

The above results seem to support earlier conjectures
\cite{thooft}
that black holes do belong to a certain class of string
theories.

\section{DISCUSSION}
\label{sec:level6}

In this work we analyzed the statistical mechanics of a gas of
black holes from the standpoint of the microcanonical ensemble.
In fact, as we showed in Section I., this is the only approach
to the problem, because the energy is not thermally distributed
among the elements of the gas.  The lack of a thermal
distribution of the energy is reflected by the non-existence of
the canonical partition function.

A gas of Schwarzschild (neutral) black holes naturally obeys the
bootstrap condition, a property related to the fact that extreme
Schwarzschild black holes are massless.  We found that quantum
corrections to the degeneracy of states play an important role
in selecting the most probable number of black holes in the gas.
The equilibrium configuration is the one for which most of the
energy is carried by a single black hole, a situation somewhat
analogous to strings.

For charged black holes such as the Reissner-Nordstr\"{o}m or its
dilaton generalization, the bootstrap condition is in general
not realized, mainly because the extreme case is not massless.
However, for such models there is a high energy ``ionization
point'' at which the gas does obey trivially the bootstrap
condition.  At such a ``critical'' point the gas is composed of
a single supermassive black hole whose size is of the order of
the Planck volume.

For all models analyzed here we find that the microcanonical
temperature is identical to the Hawking temperature of the most
massive black hole in the gas.

One motivation for the present work was the realization that the
thermodynamical interpretation of the single black hole event
horizon leads to, in our judgment, the inconsistent result of
negative canonical specific heat.  We contend that the views
presented here represent an improvement in
the understanding of black hole phenomena.  In particular, at
least viewed from the present statistical model, black hole
states can
decay into or scatter with other black hole states.  This lends
support to a completely quantum coherent view of black hole
``evaporation''.  Furthermore, arguments presented in Section V.
seem to indicate that Schwarzschild black hole scattering
amplitudes may obey a duality symmetry very similar to that of
strings.  One could then argue that black holes belong to a
special class of string theories, as conjectured
by 't Hooft \cite{thooft}.

As an example of a physical application the very early universe
can be regarded as a black hole consisting of one
very massive black hole surrounded by countless massless others,
a typical equilibrium configuration, enclosed in a small volume
of the order of the Planck volume.  It is possible that such an
inhomogeneous
initial equilibrium energy distribution gave rise to
structures and thus to galaxy formation.  Such a suggestion
was put forward by Carlitz \cite{carl} long ago in the
context of hadronic matter where an inhomogeneous
equilibrium distribution also occurs.

\section{ACKNOWLEDGMENTS}
\label{sec:level7}

We wish to thank N. Ishibashi for discussions at an early stage
of the research described in this paper.  This research was
supported in part by the U. S. Department of Energy under Grant
No. DE-FG05-84ER40141.

\end{document}